\documentclass[]{aa}
\usepackage{epsfig}
\usepackage{natbib}
\usepackage{txfonts}

\newcommand{\sub}[1]{_{\rm #1}}

\newcommand{\changed}{}
\newcommand{\newchanged}{}

\begin{document}
\title{Optimization of mapping modes for heterodyne instruments}
\author{V. Ossenkopf}
\institute{1. Physikalisches Institut der Universit\"at 
zu K\"oln, Z\"ulpicher Stra\ss{}e 77, 50937 K\"oln, Germany
\and
SRON Netherlands Institute for Space Research, P.O. Box 800, 9700 AV 
Groningen, Netherlands
\and
 Kapteyn Astronomical Institute, University of Groningen, P.O. Box 800, 
 9700 AV Groningen, Netherlands
}

\abstract
{Astronomic line mapping with single-pixel heterodyne instruments is usually
performed in an on-the-fly (OTF) or a raster-mapping mode
depending on the capabilities of the telescope and the instrument. 
The observing efficiency can be increased by combining several 
source-point integrations with a common reference measurement. This is
implemented at many telescopes, but a thorough investigation
of the optimum calibration of the modes and the best way of performing
these observations is still lacking.
}
{
We derive optimum mapping strategies and the corresponding calibration schemes 
based on the known instrumental performance in terms of system stability 
and slew times.
}
{
We use knowledge of the instrumental stability obtained by an
Allan variance measurement to derive a mathematical
formalism for optimizing the setup of mapping observations.
Special attention has to be paid to minimizing of
the impact of correlated noise introduced by the common
OFF integrations and to the correction of instrumental drifts.
Both aspects can be covered using a calibration scheme that
interpolates between two OFF measurements and an appropriate
OFF integration time. 
}
{
The total uncertainty of the calibrated data consisting of
radiometric noise and drift noise can be minimized by adjusting
the source integration time and the number of {\changed data} points observed
between two OFF measurements. It turns
out that OTF observations are very robust. They provide a low
relative noise, even if their setup deviates considerably from
the optimum. Fast data readouts are often essential to minimize
the drift contributions. In particular, continuum measurements
may be easily spoiled by instrumental drifts.
The main drawback of the described mapping modes is the
limited use of the measured data at different spatial or spectroscopic
resolutions obtained by additional rebinning. 
}
{}

\maketitle

\section{Introduction}
\label{sect_intro}

Mapping of astronomical objects with single pixel receivers 
requires a dynamical scanning of the object with the telescope,
so that different coordinates are observed at different times.
The observing scheme is complicated by the fact that all
astronomical receivers are affected by gain instabilities 
\citep[see e.g.][]{Kraus, RohlfsWilson}, so that the sensitivity
is a function of time as well. This dilemma can be solved by
regular observations of a reference position on short time scales
compared to the drift time scale. Thus all mapping schemes
include a short loop for source-reference measurements and
a longer timescale for scanning the whole 
map{\changed{}\footnote{Similar approaches are required to map
objects with array receivers that cover only a part of the object,
but there information from different pixels can be combined to
quantify drifts leading to more flexible and efficient
observing schemes \citep[see e.g.][]{Emerson, Reichertz}.}}. One can
distinguish between symmetric observing modes, where one reference
measurement is done for each source point, {\changed using equal
integration times in both phases,}  and asymmetric modes
where a reference measurement is done only after observing a number
of source map points. {\changed Examples of symmetric modes are 
dual-beam switch raster maps or frequency-switch on-the-fly maps}. 
Each point can be treated individually and
the optimum timing can be computed following the formalism
developed in \citet[][paper I]{Allanpaper}. 

Asymmetric modes are {\changed position-switch} on-the-fly (OTF) 
maps\footnote{Through the rest of the paper we will use
the term OTF map synonymous for position-switch on-the-fly maps.}
and asymmetric raster
maps. In an OTF map, the telescope continuously scans the area
to be mapped, while the detector integrates and data are read out
at a high rate. {\changed Every individual data
dump represents one point on the map, and their distance is determined
by the readout rate and the scan velocity.} After a finite
{\changed number of points, defining one scan}, the
telescope slews to a position free of emission (OFF) for the reference
measurement and then returns to the map for the next scan. The
theoretical foundations for efficient OTF mapping schemes were
laid by \citet{Mangum, Beuther}, and \citet{SchiederKramer}. Asymmetric raster
maps are similar to OTF maps except that dead times
occur when the telescope moves between different points on a map.
Because of the similarity we restrict ourselves here to 
analysis of OTF maps and discuss the differences for raster maps
only in Appendix C. {\changed \citet{Mangum} has shown that OTF 
mapping is a very efficient mode} for the observation of large fields 
in the sky with single-pixel receivers. The high efficiency results
from the continuous scanning and integration avoiding dead times 
between the observation of adjacent points and from the
reuse of the observation of a single reference position for the 
calibration of several data points. The OTF mapping imposes, however, 
harder requirements to the pointing and timing behavior of the 
telescope, which may not always be fulfilled.  

\mbox{}\citet{SchiederKramer} show that the knowledge of the 
system stability can be used to derive the optimum approach
for performing actual observations. They computed the timing 
parameters that provide the minimum uncertainty of the calibrated data,
composed of radiometric noise and drift noise, per unit observing 
time. Unfortunately, their computations were restricted to fluctuations
with an $1/f^{\alpha}$ power spectrum with a spectral index $\alpha$
of 2 and 3. When designing the mapping observing modes for
HIFI, the heterodyne instrument of the Herschel Space Observatory, 
to be launched in 2008 \citep{HIFI}, we {\changed measured, however,
a wide variety of spectral drift indices \citep{Allanpaper}.
We noticed that it is important to distinguish between
{\it spectroscopic drifts}, which characterize the variation
of spectra after a zero-order baseline subtraction, i.e. after
the correction for fluctuations that affect all channels in
the same way, and {\it total-power drifts}, which characterize
the variation of the raw spectra, thus being typically dominated by
fluctuations of the overall gain of the instrument.
Spectroscopic drifts often show values between 2 and 3, but
we also noticed that many instrumental fluctuations were dominated 
by $1/f$ noise, i.e. following a spectral index $\alpha$ close to
unity \citep{Allanpaper}. In particular total-power drifts
often show very shallow spectra. Consequently, the results
from \citet{SchiederKramer} cannot be directly applied.  This 
requires a revision} with respect to generalizing the spectral index.
Moreover, \citet{SchiederKramer} assumed a special calibration scheme for
all mapping modes where single lines in a map are combined with a
single reference measurement for calibration, although other 
calibration schemes are possible as well. {\changed We had to 
evaluate the profit from exploiting the possibility}
to scan a series of lines in alternating directions before
going to the reference position. Thus we have repeated
their computations in a more general framework allowing for
various calibration schemes and arbitrary spectral indices
resulting in general guidelines for an optimum performance and
calibration of mapping observations.

{\changed Apart from the optimization of individual OTF scans
discussed here, there exists a number of methods to reduce 
striping effects in OTF maps, in particular baseline offsets,
either by an appropriate \textit{a posteriori} data manipulation, assuming
purely linear drifts or special correlations in the observed
structures, or by a combination of multiple maps observed in
different scanning directions \citep[see e.g.][]{Steer, Emerson84, 
Maino, Ashdown}. In fact both approaches should be combined.
Here we will focus on the optimum
observational setup which minimizes drift effects from the very 
beginning, so that the measured data for all individual data
points show a high quality, independent of the number of
coverages in which an object is observed, so that it is also
applicable to mappings of bright lines which are well detected
in a single OTF map coverage. Altogether, the strategies discussed
here should be }

The outline of the paper follows our basic approach to the 
optimization problem. In Sect. 2 we introduce the properties of the
OTF mapping mode and discuss the possible ways how measured data will be
calibrated to obtain scientific data. In Sect. 3 we evaluate the different
calibration schemes with respect to their sensitivity to drift effects.
In Sect. 4 we demonstrate the application of the different calibration
schemes to actual observations performed at the KOSMA 3~m telescope.
From the best calibration scheme we optimize the exact timing of
the observations with respect to total noise in Sect. 5. Sect. 6
discussed some limitations of OTF modes and the conclusions
for the observing mode efficiencies are summarized in Sect. 7. 

\section{Introduction to OTF observations}

\subsection{The general measurement scheme}

\begin{figure}[ht]
\includegraphics[angle=0,width=\columnwidth]{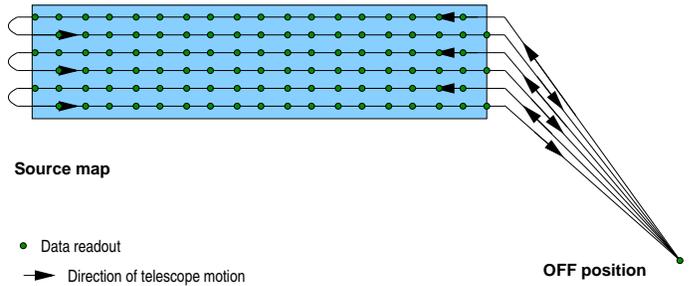}
\caption{\footnotesize {\changed Sketch} of an OTF observation.
{\changed The dots symbolize the moments when the backends
are read out. The integration starts when the telescope enters the 
rectangular area} of the map. 
In this example, the OFF position is visited after 
every two lines and the scanning direction is alternating.}
\label{fig_otf}
\end{figure}

A general introduction to OTF mapping was given by {\changed
\citet{Mangum} and \citet{Beuther}}.
The general sequence of operations is demonstrated in Fig. \ref{fig_otf}.
{\changed A map observation is split into individual scans and
their corresponding OFF measurements.
Every} scan consists of $N$ source integrations obtained while
continuously scanning the map. {\changed Each integration covers
the time $t\sub{s}$ spent between the edge of the map and the first readout
or between two subsequent readouts, symbolized by the dots 
in the picture. In the example, 
multiple lines are combined within one scan} and a turn is performed 
between subsequent lines so that they are scanned in opposite directions. 
At the end of a scan, the OFF position is visited and the reference 
measurement is performed with an integration time $t\sub{OFF}$. 
In principle, it is possible to 
go from an arbitrary position within the map to the reference position.
We will show later that it is preferable to complete an
integer number of {\changed lines in a scan}.

{\changed The measurable signal can be described by a continuous 
function $s(t)$. For each source point $i$ within a scan the
integration provides a source count rate
\begin{equation}
c_{{\rm s},i} = {1 \over t\sub{s}} 
\int_{(i-1)t_{s}}^{it_{s}} dt \; s(t)
\end{equation}
if we assign $t=0$ to the start of the scan.
Analogously, we can define the count rate for the OFF position
as the average of the signal during the OFF integration.
Because of the telescope motion during the source integration
time $t\sub{s}$, the source count
rate $c_{{\rm s},i}$ corresponds to a broadened effective
beam} along the scanning direction. \citet{Beuther} have
shown that for OTF maps where the readout is performed on a spatial
grid corresponding to a Nyquist sampling of the map, the effective
beam broadening is less than 4\,\%. They assumed a 14~dB edge taper
corresponding to a Nyquist sampling of FPBW/2.4.

However, many observations are not performed exactly on Nyquist sampling,
either by ignoring the difference between half beam width sampling and
full Nyquist sampling or by using a common sampling for different
tracers observed at different frequencies and consequently different
beam widths. In Appx. A we show that the OTF mode does not provide a
noticeable beam broadening as long as the data readout is performed  
on a time scale corresponding to a telescope motion of less than
about 0.65 HPBW. When observations ask for a lower spatial resolution,
e.g. for the comparison of line ratios, or a coarser sampling, it is 
possible to integrate longer, thus reducing the noise. Nevertheless, 
even in those cases, one should always prefer to integrate for
shorter periods, if this is technically possible, to enable
a further analysis of the data with the full resolution. {\changed
To foresee a later manual smearing to the goal resolution, the
observations then have to use a somewhat longer OFF integration
time as discussed in Sect. \ref{sect_discussion}, but this is
usually justified by the gained flexibility.}
\label{sect_genscheme}

\subsection{The calibration by a reference measurement}

The actual correction of the instrumental drift is done
by subtracting the count rate obtained during the reference
measurements at the OFF position from the count rate at 
the source points \citep{Kutner, framework}.
{\changed \begin{equation}
C_{{\rm s},i} = c_{{\rm s},i}-c\sub{OFF}
\label{eq_calib}
\end{equation}}
The radiometric noise in the calibrated data for each pixel thus
consists of noise contributions from the source and from
the OFF integrations, $\sigma\sub{noise} \propto \sqrt{1/t\sub{s}
+1/t\sub{OFF}}$. In case of no dead times it can be easily
shown that the radiometric noise {\changed for any total scan
time $N\times t\sub{s} +1t\sub{OFF}$} is minimized when using an
OFF integration time $t\sub{OFF}=\sqrt{N} t\sub{s}$, if the $N$
source integrations of one scan are calibrated with the same OFF
measurement \citep{Ball}. Although this relation is not strictly
fulfilled in the situation of non-negligible overheads, it 
remains approximately valid \citep{SchiederKramer} and is thus
widely used in current implementations of OTF observing modes at
ground-based telescopes.

{\changed However, it is not clear that the OFF measurement at the end of
an OTF scan is always the optimum reference to be
used in the subtraction. Alternatively, averages of different OFF
integrations may be used as reference. Therefore, we consider 
general case of an arbitrary reference count rate $c\sub{R}$
with an effective integration time $t\sub{R}$. 
We can distinguish three main} calibration approaches: \\
{\bf i) single OFF:} The reference position is observed for
$t\sub{R}=t\sub{OFF}$ before (or after) a series of $N$ source points
and the OFF count rate is subtracted from all source
points in the scan:
\begin{equation}
C_{{\rm s},i}=c_{{\rm s},i}-c\sub{R}
\end{equation}
where the index $i$ running from 1 to $N$ characterizes the 
different source points in a series. This approach is currently
used as standard calibration for
OTF observations at the JCMT, MOPRA and KOSMA telescopes.

{\bf ii) interpolated OFF:} The total OFF integration time
$t\sub{OFF}$ is split into two OFF observations with
half the integration time, $t\sub{R}=t\sub{OFF}/2$, 
before and after the series of $N$
source points. The reference count rate subtracted from
each source count rate is given by the linear interpolation
between the two OFF measurements
\begin{equation}
C_{{\rm s},i}=c_{{\rm s},i}-[(1-l) c\sub{R,1}+l c\sub{R,2}]\;,
\label{eq_interpolate}
\end{equation}
{\changed i.e. we use a new reference that is constructed from
two OFF measurements with the half duration.}
Here, $l$ is a time interpolation factor being $l=0$ if the 
source count rate is measured
at the time of the first OFF observation and $l=1$ if it is
measured at the time of the second OFF. It can be obtained from
\begin{equation}
l={{t\sub{R}/2+t\sub{d,1}+(i-1/2) t\sub{s}} \over {t\sub{R}
+t\sub{d,1}+t\sub{d,2}+N t\sub{s}}}
={{t\sub{R}/2+t\sub{d,1}+(i-1/2) t\sub{s}} \over {t\sub{R}
+t\sub{scan}}}
\label{eq_interpolmeasure}
\end{equation}
where the terms $t\sub{d,1}$ and $t\sub{d,2}$
stand for the dead times due to the telescope from the OFF position
to the first source point and from the last source point to the
OFF position. The number $i$ denotes the index of the source point
in the current scan and $t\sub{scan}=t\sub{d,1}+Nt\sub{s}+t\sub{d,2}$
denotes the total duration of a scan.

The actual timing of the observation can be almost identical to
case {\bf i} because the OFF measurements between the scans
are simply split into two subsequent OFF measurements with 
half duration. The only difference is that the whole observation
is bracketed between two OFF measurements with $t\sub{R}=t\sub{OFF}/2$.
This approach is currently the default setting for the OTF
calibration at the IRAM 30~m telescope.

{\bf iii) double OFF:} This approach uses the same splitting
of the OFF measurement into two parts before and after the
source series as case {\bf ii} but uses the average of both
count rates for the calibration instead of applying a linear
interpolation in time: 
\begin{equation}
C_{{\rm s},i}=c_{{\rm s},i}-\left[{1 \over 2} c\sub{R,1}+
{1 \over 2} c\sub{R,2}\right]\;.
\end{equation}
It corresponds to Eq. (\ref{eq_interpolate}) with a fixed
value $l=0.5$.
In this approach all source points in a scan are
calibrated with the fixed OFF count rate corresponding to the
value at the center of the scan in the linear interpolation.
This calibration scheme is available as optional mode at 
several ground-based telescopes.

{\changed
For all following computations we will stick to Eq. (\ref{eq_interpolate}),
which can be used for all calibration schemes when applying the 
appropriate weighting factors $l$ and $1-l$. For the case of the 
single OFF $l$ is set to 1 or 0, for the interpolated OFF Eq. 
(\ref{eq_interpolmeasure}) is used, and for the double OFF $l=1/2$.}

The obvious advantage of the linear interpolation ({\bf ii}) is
the complete cancellation of linear drifts. The
disadvantage is the production of a varying noise across
each series of source points resulting from the variable
OFF contributions. In the center of each series, where
$l=0.5$, the noise from the OFF position corresponds to an integration
time of $t\sub{OFF}$, but at the ends, where $l=0$ or $l=1$,
the noise from the OFF is higher by $\sqrt{2}$ because only a
single measurement with $t\sub{R}=t\sub{OFF}/2$ contributes.
This is actually visible in some IRAM observations where the
noise is minimal in the center of each line but increasing towards
the edges of the maps (Teyssier, priv.comm.). The effect
is relatively small for long scans with $t\sub{OFF}=\sqrt{N} t\sub{s}$ 
because the contribution from the OFF integration to the total
noise is small. Its change by a factor $\sqrt{2}$ is hardly
noticeable in most cases.
\label{sect_calibration}

\subsection{Correlated noise}
\label{sect_noisecorr}

The calibration of data from several source points with a
common OFF measurement always produces some
{\changed artificial correlation in the final maps because the
noise from the OFF measurement shows up in multiple points.
When smoothing such a map to lower resolution the noise does
not decrease with the square root of the number of points
averaged, but the noise contribution from the OFF will remain
constant. We can describe this as ``correlated noise´´ given
by the noise variance contribution from the OFF measurement
and the number of map points affected by this contribution.
From Eq. {\newchanged (\ref{eq_interpolate})} we can see that the noise variance
contribution from the OFF measurement to an individual source
point follows
\begin{equation}
\sigma\sub{R}^2 \propto { (1-l)^2 \over t\sub{R,1}} +
{ l^2 \over t\sub{R,2}}
\label{eq_corrnoise}
\end{equation}
assuming statistically independent noise in the OFF
measurements involved.}

The different ways of adding the
noise from the neighboring OFF measurements in the different
calibration approaches result in a different amount of correlated
noise throughout an OTF map. For the single-OFF calibration,
with $l=0$ or $1$ and $t\sub{R}=t\sub{OFF}$ we find a constant 
contribution from one reference measurement to all points
of a scan.

{\changed
When splitting the OFF measurement between the OTF scans into 
two parts with half the integration time, $t\sub{R}=t\sub{OFF}/2$,
the noise variance in each of these parts is twice the 
noise of the original measurement. For the double-OFF calibration
with a fixed ratio $l=1/2$ each source point
inherits a quarter of the noise variance from each of the
OFF measurements. Consequently their individual contribution falls
at half of the level from the single-OFF calibration. The sum of their 
contributions provides a noise level identical to the
single-OFF treatment. 

When the calibration involves a temporal change as for the linearly
interpolated-OFF calibration, the OFF noise will also vary from 
point to point. The linear interpolation between the two
OFF measurements bracketing a scan with $t\sub{R}=t\sub{OFF}/2$
creates an OFF noise
contribution that increases quadratically towards the boundaries 
of the scans. Compared to the single-OFF calibration the noise
sum is higher by the factor $1+4(l-1/2)^2$.
The noise variance changes within a factor two matching 
the value for the single-OFF calibration at the scan center, 
but being two times as high at the boundaries.

We can quantify the impact of this OFF noise as
correlated noise within the map by the simple parameter
$\sigma\sub{combined}^2$, given by the product of the number of pixels
showing the same noise contribution and the variance of this noise. 
This definition reflects the visual effect of correlated noise in a 
map where the eye automatically tends to integrate over parts of the
map to detect structures. The product of variance and pixel number 
corresponds to this integration. For all calibration schemes
discussed so far, the correlated noise is restricted to single
scans and can be approximated by integrating the individual noise
contributions in Eq. (\ref{eq_corrnoise}) over the scan length
ignoring the discretization of the scan in terms of source points.

The single-OFF calibration and the double-OFF calibration produce
the same correlated noise sum, as each point of a scan is treated 
with the same OFF noise variance in both cases.
For the interpolated-OFF calibration, however, we find a
correlated noise sum, $\sigma\sub{combined}^2$, which amounts
to only 2/3 of the value in the single-OFF calibration,
because the contributions from the individual noise
measurements vary quadratically across a scan (Eq. \ref{eq_corrnoise}).
Although the average total noise sum amounts to 4/3 of the value 
from the single-OFF calibration we have a lower correlated
noise. The linear interpolation thus shows a slightly
increased total noise but a decreased correlated noise
contribution compared to the single-OFF scheme.}

An obvious further step towards an increase of the efficiency
of the observations is the reuse of an OFF measurement
for two two adjacent scans so that the OFF integration
time can be reduced by a factor two. When using the full OFF
integration time for the reference time, i.e. $t\sub{R}=t\sub{OFF}$,
all the equations from Sect. \ref{sect_calibration} are still
valid. {\changed The total noise variance is not changed, but we create
additional correlations between the noise in different pixels.
The correlated noise sum is increased by a factor two, because 
the noise from one OFF measurement is spread across the two 
adjacent scans.} For the interpolated-OFF calibration with a reuse of the
OFF data for both adjacent scans the correlated noise sum 
and the average total noise variances are larger by a factor 
4/3 relative to the value obtained in the single-OFF calibration. 
Vice versa we can use an OFF calibration time of 2/3 of the
time used in the single-OFF calibration to obtain the
same amount of total noise and correlated noise in the
interpolated-OFF calibration.

{\changed
With respect to the radiometric noise we have thus the situation
that single-OFF and double-OFF calibration produce the same total
noise contribution and the same correlated noise contribution
from the OFF measurement to each source measurement.For the double-OFF
integration we could reduce the OFF integration time by a factor
two still maintaining the same total noise contribution, but at 
cost of a higher correlated noise. In the 
interpolated-OFF calibration we need only an OFF integration 
time of 2/3 compared to the single-OFF calibration to obtain 
the same noise values, but on top of that we achieve a complete
cancellation of all linear drift errors.}

After these general considerations we will actually compute
the error in the calibrated data due to both instrumental
drift and the radiometric noise both from the source points
and from the OFF subtraction in the next section.
\label{sect_correlated}

\section{The data uncertainty due to noise and drift}
\label{ref_theory}

\subsection{Quantitative estimate}

The total uncertainty of the measured data is given by the
sum of the uncertainties from radiometric noise and instrumental
drifts. With known fluctuation spectra $S(f) \propto 1/f^{\alpha}$
for the two contributions the total data uncertainty can be
computed as demonstrated by \citet{SchiederKramer}. They performed
the estimate for the special case of a single-OFF calibration
and spectral indices of the instrumental fluctuations $\alpha=2, 3$.
Here, we repeat these computations for the general case.

{\changed If we write the expression for the calibrated data
(Eq. \ref{eq_calib}) for an arbitrary start time $t$ and the 
general calibration approach expressed by the weighting factor $l$
we obtain}
\begin{eqnarray}
C_{{\rm s},i}&=&c_{{\rm s},i}(t)-(1-l) c\sub{R,1}(t)-l c\sub{R,2}(t)
\\
&=&{1 \over t\sub{s}} 
\int_{t+t\sub{R}+t\sub{D,1}}^{t+t\sub{R}+t\sub{D,1}+t\sub{s}} dt' \; s(t')
\nonumber \\
&&- (1-l) {1 \over t \sub{R}} \int_{t}^{t+t\sub{R}} dt' \; s(t') 
- l {1 \over t\sub{R}} 
\int_{t+t\sub{scan}+t\sub{R}}^{t+t\sub{scan}+2t\sub{R}} dt' \; s(t') \;.
\end{eqnarray}
To abbreviate the notation we use here the total delay time
before a given source measurement $i$, $t\sub{D,1}=t\sub{d,1}
+(i-1)t\sub{s}$. In the same way we define $t\sub{D,2}$ as the
total delay time after a given source measurement $i$, 
$t\sub{D,2}=t\sub{scan}-t\sub{D,1}-t\sub{s}=t\sub{d,2}+(N-i)t\sub{s}$.
With the appropriate weighting factors $l$ and $1-l$,
the equation can be used for all calibration schemes discussed above.
In all schemes where the OFF measurement is split
into two separate contributions we use $t\sub{R}=t\sub{OFF}/2$,
otherwise $t\sub{R}=t\sub{OFF}$. 

{\changed Assuming ergodicity we can obtain the average total 
uncertainty of the count rate from a time-average}
\begin{equation}
\sigma_{C}^2(i)=
\left\langle ( C_{{\rm s},i} - \langle C_{{\rm s},i} \rangle_t )^2
\right\rangle_t
\label{eq_sigma_basic}
\end{equation}
where we treat the measurement as a continuous function, ignoring
that it is performed only in discrete steps.

It can be easily seen that the maximum uncertainty occurs for
weak signals where the count rates on the source and on the
OFF position basically are the same. Thus we consider the
worst case assuming $\langle c_{{\rm s},i}(t) -
(1-l) c\sub{R,1}(t)-l c\sub{R,2}(t) \rangle_t=0$.
Then the second term in Eq. (\ref{eq_sigma_basic}) vanishes and
we can rewrite it as
\begin{eqnarray}
\sigma_{C}^2(i)&=&
\left\langle c_{{\rm s},i}(t)^2 \right\rangle_t
+ (1-l)^2 \left\langle c\sub{R,1}(t)^2 \right\rangle_t
+ l^2 \left\langle c\sub{R,2}(t)^2 \right\rangle_t \nonumber \\
&&-2 (1-l) \left\langle c_{{\rm s},i}(t)c\sub{R,1}(t) \right\rangle_t
-2 l \left\langle c_{{\rm s},i}(t)c\sub{R,2}(t) \right\rangle_t \nonumber\\
&& +2 l(1-l) \left\langle c\sub{R,1}(t)c\sub{R,2}(t) \right\rangle_t \;.
\label{eq_sigma_full}
\end{eqnarray}
The first three contributions contain the variation representing
the noise within each of the three measurements involved. The
other terms represent the cross correlation between them containing
the mutual drift terms.

The computation of all six terms follows the same approach.
We demonstrate it here only for $\left\langle c_{{\rm s},i}(t)c\sub{R,1}(t) \right\rangle_t$: 
\begin{equation}
\left\langle c_{{\rm s},i}(t)c\sub{R,1}(t) \right\rangle_t =
{1 \over t\sub{s} t \sub{R}} \left\langle 
\int_{t}^{t+t\sub{R}} dt' \int_{t+t\sub{R}+t\sub{D,1}}^{t+t\sub{R}+t\sub{D,1}+t\sub{s}} dt'' \; s(t') s(t'') 
\right\rangle_t \;.
\end{equation}
With {\changed the coordinate transformation to time variables 
$\tau'=t+t\sub{R}-t'$ and $\tau''=t+t\sub{R}-t''$} we obtain
\begin{eqnarray}
\left\langle c_{{\rm s},i}(t)c\sub{R,1}(t) \right\rangle_t \!&=\!&
{1 \over t\sub{s} t\sub{R}}
\\ &\times\!&  
\int_{0}^{t\sub{R}} d\tau' \int_{0}^{t\sub{s}} d\tau'' \left\langle
s(t-\tau') s(t-\tau''+t\sub{D,1}+t\sub{s}) 
\right\rangle_t \nonumber
\end{eqnarray}
where we have also exchanged the sequence of integration and 
time-averaging.
The integrals can be evaluated using the auto-correlation function
of the fluctuation spectrum. For a power-law noise spectrum $S(f)
\propto 1/f^{\alpha}$ with spectral indices {\changed $0< \alpha \le 3$}
the auto-correlation function can be evaluated 
as\footnote{For $\alpha=1$ there exists a logarithmic deviation
so that equation does not hold for this particular value.}
\begin{eqnarray}
\gamma(\tau)&=&\langle s(t+\tau) s(t) \rangle_t \nonumber\\
&=&g_0 -g_{\alpha} |\tau|^{\alpha-1}
\end{eqnarray}
assuming zero averages \citep{SchiederKramer}. {\changed This
relation and the properties of this kind of autocorrelation
functions are extensively used in studies of fractal
and turbulent processes \citep[e.g.][]{Peitgen, Bunde, Frisch}.

Exploiting this relation, the integration can be carried out
as demonstrated in Appendix B resulting in}
\begin{eqnarray}
\left\langle c_{{\rm s},i}(t)c\sub{R,1}(t) \right\rangle_t &=&
g_0-{g_{\alpha} \over \alpha(\alpha+1) t\sub{s} t \sub{R}}
\left\{[t\sub{D,1}+t\sub{s}+t\sub{R}]^{\alpha+1}
 \right. \\
&& \qquad \left. - [t\sub{D,1}+t\sub{R}]^{\alpha+1}
- [t\sub{D,1}+t\sub{s}]^{\alpha+1}
+ t\sub{D,1}^{\alpha+1} \right\} \nonumber
\end{eqnarray}

{\changed
We can perform the integration for all terms in Eq. (\ref{eq_sigma_full})
and obtain}
\begin{eqnarray}
\label{eq_correlation}
\lefteqn{\sigma_{C,\alpha}^2(i)=
{- 2 g_{\alpha} \over \alpha (\alpha+1)} \bigg\{ t\sub{s}^{\alpha-1}
+ (1-2l+2l^2) t\sub{R}^{\alpha-1} } \\
&& + l(1-l) {(2t\sub{R}+t\sub{scan})^{\alpha+1}
-2(t\sub{R}+t\sub{scan})^{\alpha+1}
+t\sub{scan}^{\alpha+1} \over t\sub{R}^2} \nonumber\\
&& - (1-l) {(t\sub{R}+t\sub{D,1}+t\sub{s})^{\alpha+1}
\!-\!(t\sub{R}+t\sub{D,1})^{\alpha+1}
\!-\!(t\sub{D,1}+t\sub{s})^{\alpha+1}
\!+\!t\sub{D,1}^{\alpha+1} \over t\sub{R}t\sub{s}} \nonumber\\
&& \left.- l {(t\sub{R}+t\sub{D,2}+t\sub{s})^{\alpha+1}
-(t\sub{R}+t\sub{D,2})^{\alpha+1}
-(t\sub{D,2}+t\sub{s})^{\alpha+1}
+t\sub{D,2}^{\alpha+1} \over t\sub{R}t\sub{s}}
\right\} \nonumber \;.
\end{eqnarray}
The first two terms contain the fluctuations within the source
and the OFF measurements {\changed (Eq. \ref{eq_alpharadio})}, 
the third term represents the drift
between the two involved OFF measurements and the last two terms
characterize the drift between the source measurement and the
two OFF measurements.

The coefficient $g_{\alpha}$ giving the amplitude of the fluctuations
can be determined by an Allan variance measurement \citep{Allan}. The Allan
variance measures the variance of the difference of the signal
between subsequent intervals in a long time series of data dumps
as a function of the length of the intervals. A comprehensive 
introduction to this technique was given in paper I.
The Allan variance spectrum can be computed in the same way
as laid out above assuming zero delays\footnote{The original 
definition of the Allan variance by \citet{Allan} is lower by 
the factor 1/2.} {\changed For $l=0$, $t\sub{s}=t\sub{R}=t\sub{bin}$, and
{\newchanged $t\sub{D,2}=t\sub{D,2}=0$ we can transform Eq. (\ref{eq_correlation})}
into}
\begin{equation}
\sigma\sub{A,\alpha}^2={2 g_{\alpha} \over \alpha (\alpha+1)} \;
4(2^{\alpha-1}-1) t\sub{bin}^{\alpha-1}
\end{equation}
where $t\sub{bin}$ denotes the length of the data intervals.
This allows us to express the uncertainty of the calibrated OTF data
in terms of the Allan variance spectrum.

However, the fluctuations of any signal are not only characterized by
a single power spectrum but they consist at least of a superposition
of white noise with a spectral index $\alpha=0$ and an instrumental
drift contribution with some steeper spectral index $\alpha$. Fortunately,
we expect no correlation between the radiometric white noise
and the instrumental drift, so that both the Allan variance spectrum
and the uncertainty of the calibrated data from an OTF observation
simply are {\changed the sum of both contributions, $\sigma\sub{A}^2=
\sigma\sub{A,0}^2+\sigma\sub{A,\alpha}^2$ and 
$\sigma\sub{C}^2=\sigma\sub{C,0}^2+\sigma\sub{C,\alpha}^2$.} The Allan 
variance of the white noise contribution is given by
\begin{equation}
\sigma\sub{A,0}^2={2 \langle s(t) \rangle_t^2 \over B\sub{Fl} t\sub{bin}}
\end{equation}
{\changed and the white noise contribution to the OTF measurement is}
\begin{equation}
\sigma\sub{C,0}^2={\langle s(t) \rangle_t^2 \over B\sub{Fl}}
\left( {1 \over t\sub{s}} + {1-2l+2l^2) \over t\sub{R}} \right)
\end{equation}
where $B\sub{Fl}$ denotes the fluctuation bandwidth of the
radiometric noise. 

{\changed
With the definition of the Allan time $t\sub{A}$ as the
bin size $t\sub{bin}$ where the drift contribution and the radiometric 
noise in the measured Allan variance spectrum show the same
magnitude (paper I)}, we can relate the radiometric noise
to the coefficient $g_{\alpha}$. We
obtain the coefficient of the drift contribution as
\begin{equation}
g_{\alpha}={ \alpha (\alpha+1) \langle s(t) \rangle_t^2 \over
4 (2^{\alpha-1}-1) B\sub{Fl} t\sub{A}^\alpha} \;.
\label{eq_g_alpha}
\end{equation}

Finally we can compare the total uncertainty of the
calibrated data $\sigma_{C}^2(i)$ to the unavoidable
uncertainty due to the radiometric noise in an equivalent
measurement with an ideal instrument without any drifts, 
in an ideal observation without the need for an OFF measurement.
If we assume that this observation uses the total observing
time for the $N$ points of an OTF cycle in a given map,
$t\sub{tot}=t\sub{OFF}+t\sub{scan}$, the resulting 
data uncertainty is
\begin{equation}
\sigma_{C,{\rm ideal}}^2={N \langle s(t) \rangle_t^2 \over B\sub{Fl} 
t\sub{tot}} \;.
\label{eq_sigma_ideal}
\end{equation}

{\changed When we combine Eqs. (\ref{eq_g_alpha}) and (\ref{eq_sigma_ideal})
to substitute $g_\alpha$, add the radiometric and drift noise contributions
$\sigma\sub{C,0}^2$ and $\sigma\sub{C,\alpha}^2$ and normalize the
resulting noise of the real OTF observation}
relative to the limiting ideal observation we obtain a measure for
the actual impact of all instrumental effects on the data quality
\begin{eqnarray}
\label{eq_totaluncert}
\lefteqn{{\sigma_{C}^2(i) \over \sigma_{C,{\rm ideal}}^2}
= {x\sub{tot} \over N}
\left\{ { 1 \over x\sub{s}} + {1-2l+2l^2 \over x\sub{R}}
\right. } \\
&& -{ 2 \over 4 (2^{\alpha-1} -1)}
\bigg[ x\sub{s}^{\alpha-1}
+ (1-2l+2l^2) x\sub{R}^{\alpha-1}  \nonumber \\
&& + l(1-l) {(2x\sub{R}\!+\!x\sub{scan})^{\alpha+1}
\!-\!2(x\sub{R}\!+\!x\sub{scan})^{\alpha+1}
\!+\!x\sub{scan}^{\alpha+1} \over x\sub{R}^2} \nonumber\\
&& - (1-l) {(x\sub{R}\!+\!x\sub{D,1}\!+\!x\sub{s})^{\alpha+1}
\!-\!(x\sub{R}\!+\!x\sub{D,1})^{\alpha+1}
\!-\!(x\sub{D,1}\!+\!x\sub{s})^{\alpha+1}
\!+\!x\sub{D,1}^{\alpha+1} \over x\sub{R}x\sub{s}} \nonumber\\
&& \left.\left. - l {(x\sub{R}\!+\!x\sub{D,2}\!+\!x\sub{s})^{\alpha+1}
\!-\!(x\sub{R}\!\!+\!x\sub{D,2})^{\alpha+1}
\!-\!(x\sub{D,2}\!+\!x\sub{s})^{\alpha+1}
\!+\!x\sub{D,2}^{\alpha+1} \over x\sub{R}x\sub{s}}
\right] \right\} \nonumber
\end{eqnarray}
where we have transformed all time scales relative to the Allan time
$t\sub{A}$ with $x\sub{tot}=t\sub{tot}/t\sub{A}$,
$x\sub{s}=t\sub{s}/t\sub{A}$ and so on.

We find two essential contributions: the first two terms
characterize the radiometric noise of the observations. This noise
is higher than the radiometric noise in the ideal observation due
to the $x\sub{R}$-term containing the noise from the OFF measurement.
Without this term the radiometric noise ratio would be unity. All terms in
the brackets characterize the drift contribution to the
total data uncertainty. The different terms stand here for the
drift occurring during the different time lags involved in the
measurement. The ratio between the drift noise and the radiometric
noise of the observed data can be computed by simply dividing these
two contributions.
\label{sect_driftnoisecomputation}

\subsection{Comparison of the different calibration schemes}

With Eq. (\ref{eq_totaluncert}) we can draw
quantitative conclusions on the different calibration schemes.
We have computed the data uncertainty $\sigma_{C}^2(i)/\sigma_{C,{\rm ideal}}^2$
as a function of the scan length $N$, the spectral index of the 
instrumental drift $\alpha$, the position of a source
point within the OTF scan i, the source point integration
time $x\sub{s}$, and the dead times between the
OFF measurement and the source integrations in the {\changed scan}.
To avoid too many parameters in the following examples we simplify them by
assuming that the two dead times for moving from the source to the
OFF position and vice versa are the same, $x\sub{d,1}=x\sub{d,2}$.
This is well fulfilled for most observations with the Herschel satellite
and still a reasonable approximation for most ground-based telescopes.
Moreover, we assume in this section that the total integration
time on the OFF position follows the standard rule
$x\sub{OFF}=\sqrt{N} x\sub{s}$ derived for an {\changed ideal telescope}.

\begin{figure}[ht]
\includegraphics[angle=90,width=8.5cm]{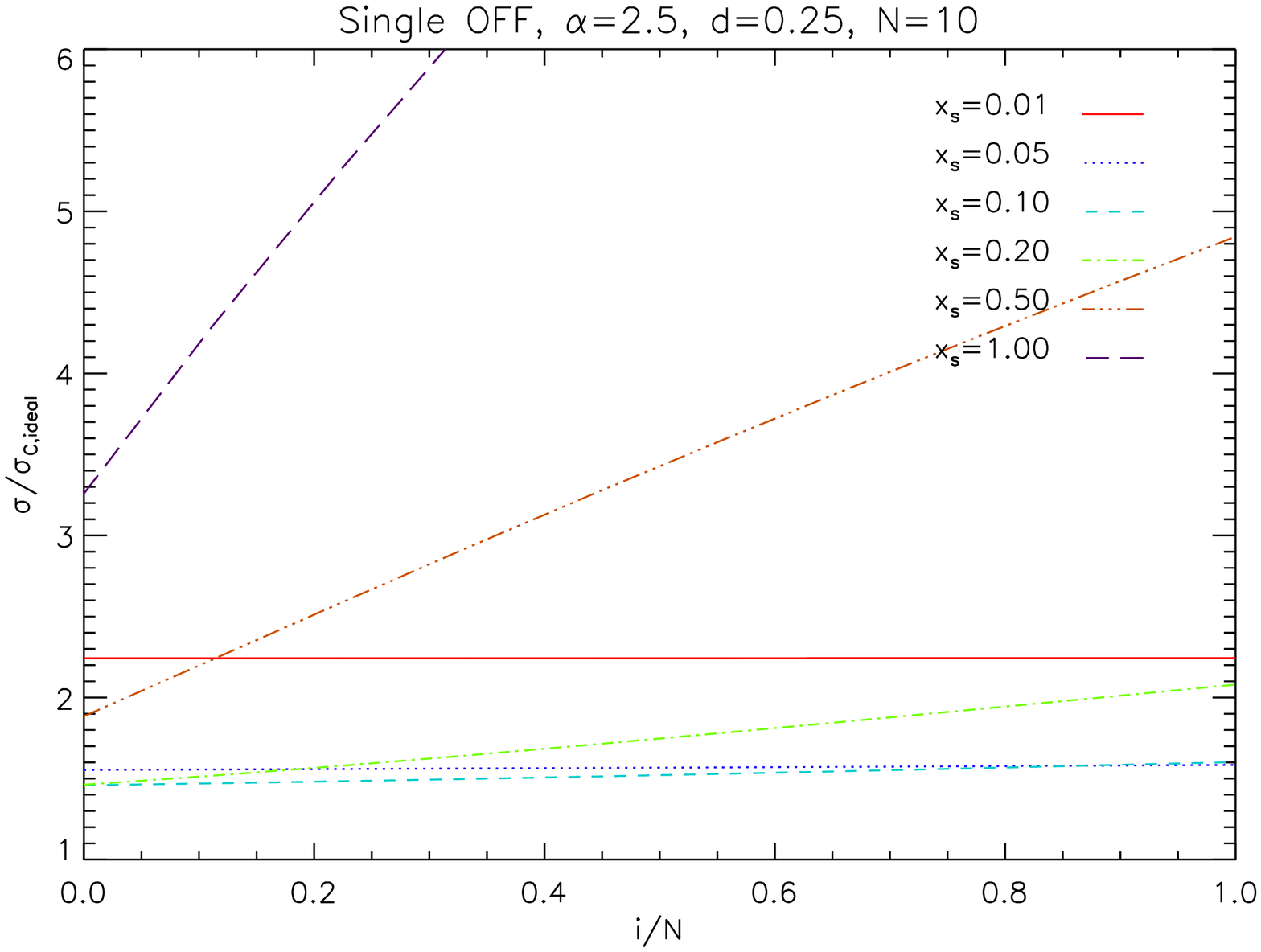}\\
\includegraphics[angle=90,width=8.5cm]{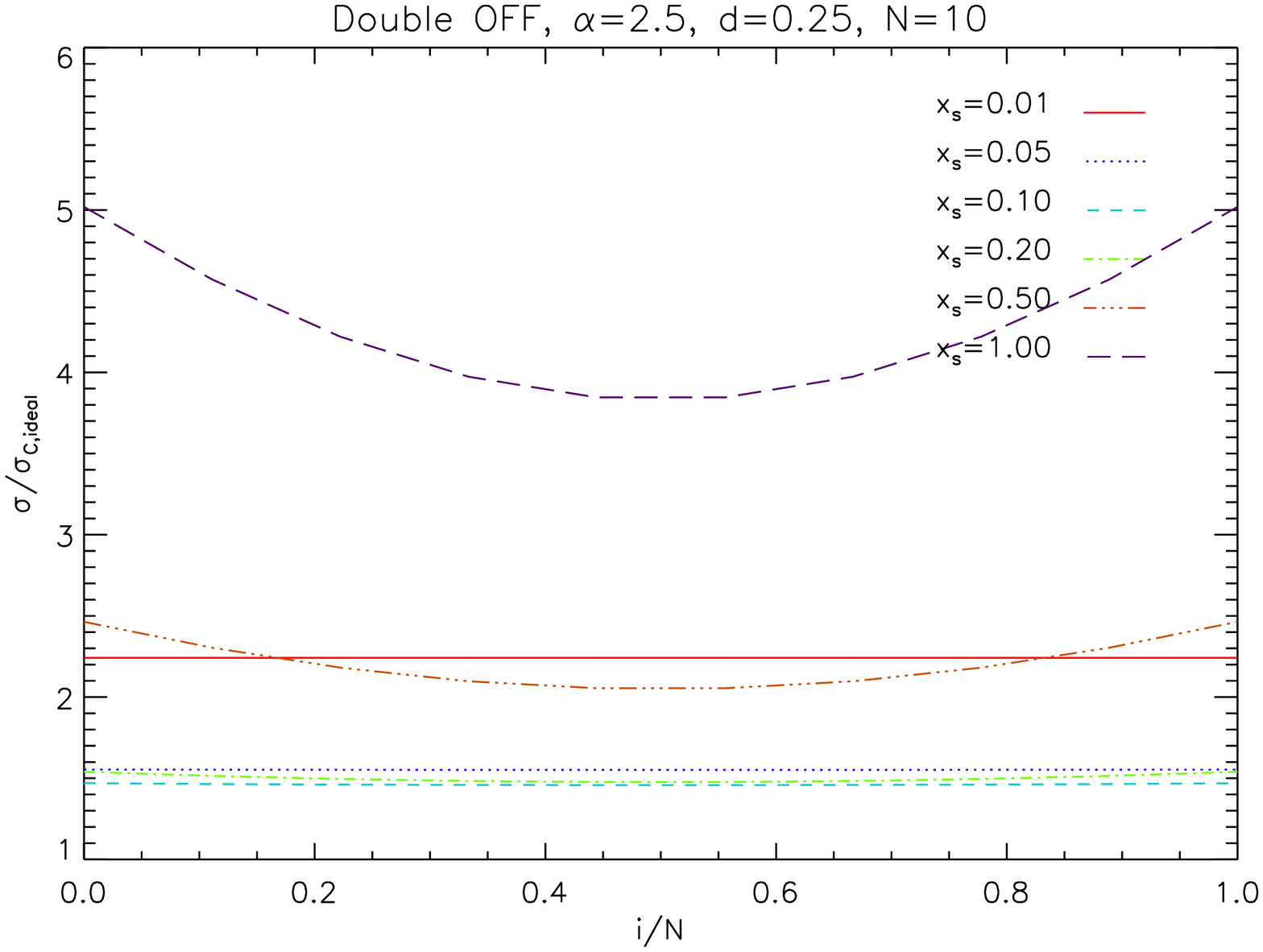}\\
\includegraphics[angle=90,width=8.5cm]{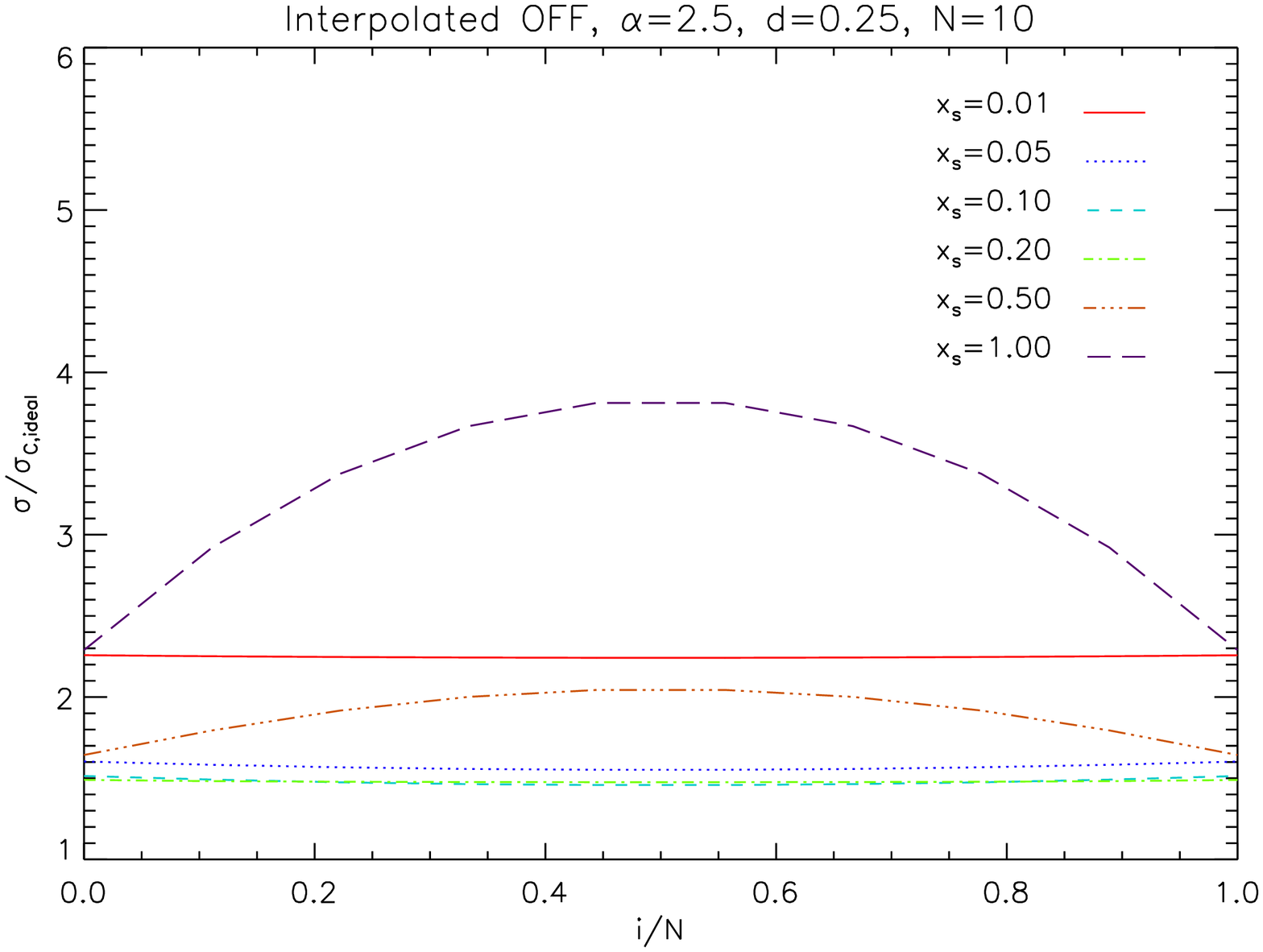}\vspace{-0.6cm}\\
\unitlength1cm
\begin{picture}(8.6,0.1)
\put( 1.2,18.2){\bf a)}
\put( 1.2,11.8){\bf b)}
\put( 1.2,5.4){\bf c)}
\end{picture}
\caption{\footnotesize Variation of the total data uncertainty across an OTF {\changed scan}
obtained in the different OTF calibration schemes.
The RMS of the fluctuations is plotted relative to the RMS
which would be obtained by an ideal instrument in the same total time.
Part {\bf a} shows the result from the single-OFF calibration with
the OFF measured before the scan, {\bf b} the double-OFF calibration, and 
{\bf c} the interpolated-OFF calibration. A scan length
$N=10$, a spectral index $\alpha=2.5$, and a total dead time
$x\sub{d,1}+x\sub{d,2}=0.25$ were used.}
\label{fig_scancompare}
\end{figure}

In Fig. \ref{fig_scancompare} we compare the three standard
calibration schemes from Sect. \ref{sect_calibration} for an
example scan consisting of $N=10$ points. In the simulation
a spectral index of the instrumental drift $\alpha=2.5$
was used, typical of spectroscopic fluctuations
\citep{SchiederKramer,Allanpaper}, and the total dead time given as
the sum of the dead times before and after an OFF measurement
was assumed to be a quarter of the Allan time which is a typical
value for many Herschel observations. The figure shows the 
normalized total noise RMS as a
function of the position of a source point within the {\changed scan}
for different source integration times $x\sub{s}$.

For all three cases we find that the shortest integration
time, $x\sub{s}=0.01$, results in a relatively
high noise. This can be easily understood by the low efficiency
of this observation where only a short integration time 
is spent on the source but a large fraction of the total
cycle is occupied by the dead times. In this relatively
fast cycle no instrumental drifts are seen and the noise is 
barely varying across the {\changed scan}. The variation of the radiometric
noise across the scan by up to 11\,\% for a 10-points scan in the
interpolated-OFF calibration scheme discussed in 
Sect. \ref{sect_calibration} actually is much lower 
because all 10 points are measured close to the center of
the time interval between the two OFF measurements.

In contrast, we find a strong noise variation across the
scan for all calibration schemes when the source 
integration time per point is in the order of the Allan time.
All data are dominated by instrumental drifts also leading
to a high total noise. An intermediate integration time
per source point results in the lowest overall data uncertainty.
In this example the optimum falls between about 0.1 Allan times
for the single-OFF calibration and 0.2 Allan times for the
interpolated-OFF calibration. Further parameter studies show 
that the optimum falls at shorter integration times for
shorter dead times and for larger numbers of source points in a scan.

Comparing the calibration schemes shows extreme differences
in the impact of the instrumental drift. The single-OFF
calibration has a huge sensitivity to instrumental drifts.
{\changed When using the OFF measurement before the scan for calibration, 
as shown in the figure, the total data uncertainty grows
monotonically towards the end of the scan, for $x_s=1$ it
exceeds even the selected plot range going up to a value of 11.}
For the corresponding scheme using the OFF after the {\changed scan},
the plot would be mirrored.
The advantage of the interpolated-OFF calibration relative
to the double-OFF calibration is also clearly visible. The
latter has a much larger data uncertainty at the ends of
the scan due to instrumental drifts. Compared to this uncertainty,
the variation of the radiometric noise from the OFF calibration 
in the interpolated-OFF scheme, well noticeable only for 
$x\sub{s}=0.05$, is a very small contribution.
In the center of the scan double-OFF and interpolated-OFF
calibration necessarily have to agree. For all cases with
a noticeable instrumental drift, the interpolated-OFF calibration is
superior to the double-OFF calibration. The latter one is
slightly better for fast scans where instrumental drifts play no role.

{\changed When performing the same computations for longer OTF 
scans, which are usually used when mapping large areas on the sky 
with ground-based telescopes, we find the same qualitative behavior
as shown in Fig. \ref{fig_scancompare}, but an increase of all drift
effects, due to the longer times covered, and a further reduction 
of the radiometric noise variation across the scan, so that this
turns invisible in the corresponding plots.}
An extended parameter study has shown that the data uncertainty
due to drift effects grows with the spectral index of the
fluctuation spectrum $\alpha$ and with the dead times before
and after the OFF measurements. {\changed For $1/f$ spectra,
we obtain a very weak dependence of the maximum drift noise
on the integration time, the scan length or the dead time.
For moderate scan lengths, $\la 100$ points, and integration times
covering a noticeable fraction of the Allan time, the maximum total
noise RMS is always approximately twice the ideal noise RMS.
However, $1/f$ spectra are often also correlated with very short 
Allan times, then limiting the observations. One has to be aware 
that all time scales have to be considered relative to the Allan time. 
For steeper noise spectra, the drift provides the main limitation 
to the possible scan lengths.
Dead time, Allan time and drift index are determined
by instrument and telescope so that their design should be 
directed towards a minimization. 
The main prerequisite for any accurate mapping observation is a
a low instrumental drift expressed by a long Allan time and/or
a shallow drift index.}

One can still improve the observing efficiency by an appropriate
calibration scheme and an optimized setup of the observations.
The drift uncertainty is increased by a larger number of source points
in each scan but reduced in the case of shorter integration times
per source point. The mutual optimization of these two parameters
leads in general to the smallest uncertainties for scans with
a large number of points but very short integration times. This
may be limited, of course, by the size of the region to be mapped and
data rate which can be taken with the instrument. A detailed
optimization taking both effects into account is given in Sect.
\ref{sect_optimize}. 

Looking at the overall pictures it is clear that the interpolated-OFF
approach is in general the most robust one. The 
single-OFF calibration is easily disqualified compared to the other 
two schemes. {\changed The double-OFF calibration can be slightly 
better than the interpolated-OFF calibration if the integration time
is very short and we have an accurate knowledge of the instrumental
drift behavior. Taking the usual uncertainty and the statistical
fluctuations of the actual drift behavior into account, leads us,
however, to the general preference of the interpolation scheme.
These results hold independent of
the possible split or combination of the OFF integrations with
respect to the neighboring scans (Sect. \ref{sect_noisecorr}) as
this would only affect the correlated noise sum, not changing
the noise amplitude computed here.}

\section{Application to observed data}

The different calibration schemes were tested using existing 
molecular line observations performed with the KOSMA 3~m
telescope. An arbitrary OTF patch was taken from a larger survey
$^{13}$CO 2-1 survey of the Cygnus X region \citep{Schneider}.
The observations were taken in the ordinary OTF mode where after
each line of the patch, containing 20 source integrations of 5~s,
one OFF integration of 23~s, corresponding to $\sqrt{20} \times 5$~s,
was performed. For the slew from the end of an OTF scan
to the OFF position a dead time of 19~s was needed, the slew
from the OFF position to the beginning of the subsequent OTF scan
took 12~s.
The spectral resolution of the used backend is 360~kHz
and the corresponding fluctuation bandwidth 560~kHz.
The spectroscopic Allan time of the instrument at this 
resolution is about 120~s. The Allan time of the whole system
including the atmosphere is estimated to be approximately 80~s. 
The drift index of the fluctuations
falls between 2 and 3. For all computations we assume 2.5 here.
The single-sideband system temperature during the observations
was about 350~K.

To emphasize the drift effects we first consider maps of line
integrated intensities, where the full velocity range of the
$^{13}$CO line from 4 to 8 km/s was integrated corresponding
to an effective bin width of 2.9~MHz. As the binning reduces
the radiometric noise of the data, this corresponds to a
reduction of the Allan time, where radiometric and drift noise
have equal amplitudes. Following the formalism developed in 
\citet{Allanpaper} we can compute an effective Allan time at
2.9~MHz bin width of about 30~s. One OTF cycle corresponds to 
approximately five Allan times at this resolution so that
we expect to notice drift effects in the integrated maps.

\begin{figure*}[ht]
\includegraphics[angle=90,width=15.3cm]{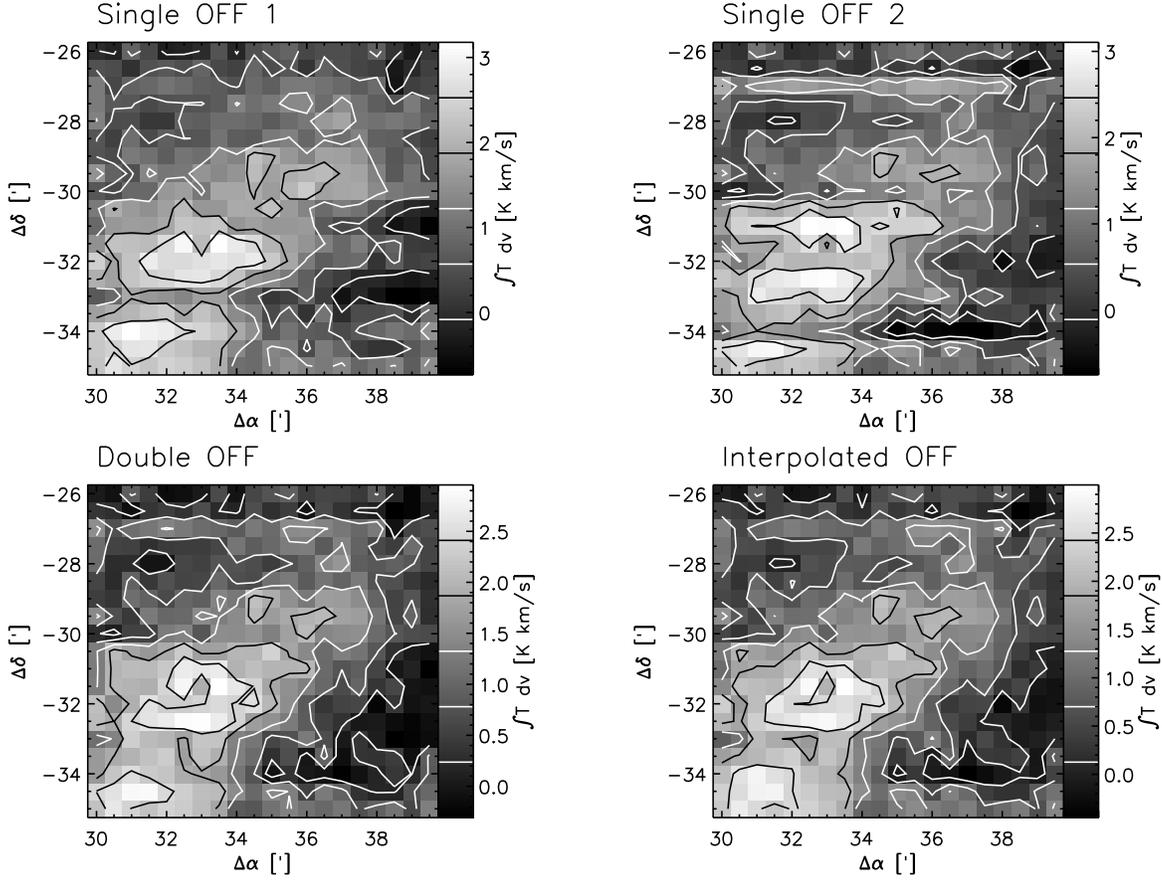}
\caption{\footnotesize Demonstration of the influence of the different calibration
schemes on the appearance of the produced line maps. The maps in the
upper panels were obtained when calibrating with the single OFF
measurement before and after each line. The lower left panel shows
the result with a fixed sum from the two adjacent OFF measurements
($l=1/2$) and the lower right panel shows the result when using
a time interpolation between both OFF measurements. The
map was measured in horizontal stripes. After each line an OFF measurement
was taken.
}
\label{fig_observedlinemap}
\end{figure*}

Figure \ref{fig_observedlinemap} shows the integrated line
maps obtained in the different calibration schemes. 
We can compare the observed structure  with the noise
computed from Eq. (\ref{eq_totaluncert}). An ideal observation, 
{\changed spending all the observing time for the map
integration}, would result in a radiometric noise of $\sigma\approx 0.1$~K
{\changed (see Eq. \ref{eq_sigma_ideal}).}
Due to overheads and the noise contribution from the OFF 
measurement, the actual radiometric noise is higher by a factor
between 1.31, obtained for the double-OFF calibration and in the
scan center for the interpolated-OFF calibration, and 1.37
for the single-OFF calibration. The drift noise 
$\sigma_{C,{\rm drift}}$ varies across the 20 points of the 
scan between 0.65 and 0.72 $\sigma_{C,{\rm rad.}}$ for the 
double-OFF calibration, between 0.56 and 0.65 $\sigma_{C,{\rm rad.}}$
for the interpolated-OFF calibration, between 0.77 and 1.45 
$\sigma_{C,{\rm rad.}}$ for the single-OFF calibration using 
the OFF before the scan, and between 0.82 and 1.49
$\sigma_{C,{\rm rad.}}$ for the single-OFF calibration using the 
OFF after the scan. While the drift noise should be hidden in
the radiometric noise for the double-OFF and the interpolated-OFF 
calibrated data, it should be clearly noticeable
in the single-OFF calibrated maps towards the ends of the scans
which are most apart from the corresponding OFF measurement.

\begin{figure}[ht]
\includegraphics[angle=90,width=\columnwidth]{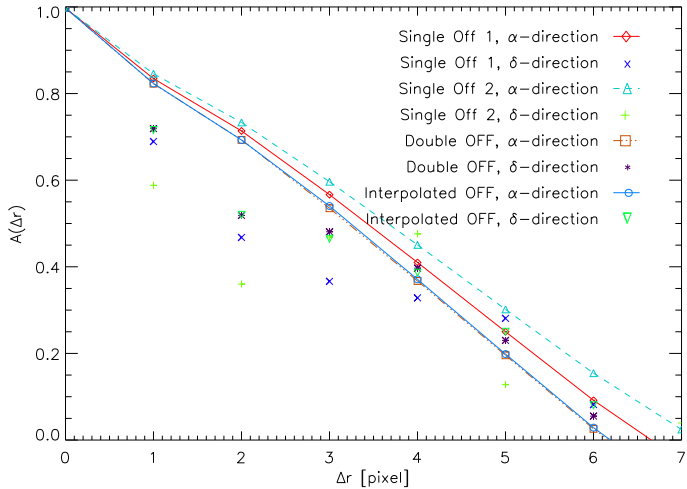}
\caption{\footnotesize Autocorrelation function for the maps from Fig. 
\ref{fig_observedlinemap} measured in the two perpendicular directions
$\alpha$ and $\delta$. {\changed To emphasize the scan direction $\alpha$,
the symbols for that direction are connected by lines.}
For an isotropic structure 
both directions should show the same values.}
\label{fig_acflinemap}
\end{figure}

We can clearly recognize the very ``stripy'' structure in
the two single-OFF calibration maps, but the stripes are not 
restricted to one end of the scans but cover large parts of the 
map. The drift effects are visible as global
structures in the integrated line maps, but for each single source
point they could still be hidden in the radiometric noise.
We have to keep in mind that the Allan variance is only a
statistical measure to characterize the drift behavior. Thus we cannot
expect to find uniform drift effects in all scans but we will always
find lines with stronger and weaker indications of instabilities. 
Eq. (\ref{eq_totaluncert}) only gives the $1\sigma$
uncertainties of a stochastic process. 

We can quantify the 
``stripiness'' of the maps by comparing variations in the map
in the direction of the OTF scans and perpendicular to them
assuming that the observed astrophysical structure will be more or
less isotropic. Figure \ref{fig_acflinemap} shows the autocorrelation
functions $A(\vec{\Delta r}) = \langle C(\vec{r}) 
C(\vec{r}+\vec{\Delta r}) \rangle_{\vec{r}} /
\langle C(\vec{r})^2 \rangle_{\vec{r}}$ when using $\vec{\Delta r}$ parallel
and perpendicular to the scan direction for all four calibrated maps.
For an isotropic structure the autocorrelation function should decay
in both directions in the same way. We notice, however, significantly
lower values of the autocorrelation function measured in the 
$\delta$-direction for all four maps at shifts of one or two pixels
indicating variations due to the mapping structure. The effect is
largest for the single-OFF map using the OFF after the {\changed scan}
with the strongest stripes also visible by eye in Fig. 
\ref{fig_observedlinemap}.  The maps resulting from the double-OFF
calibration and the interpolated-OFF calibration have almost the
{\changed same, still significant, anisotropy which is, however, reduced
by a factor two compared to the single-OFF map with the OFF after the scan.}

\begin{figure*}[ht]
\includegraphics[angle=90,width=15.3cm]{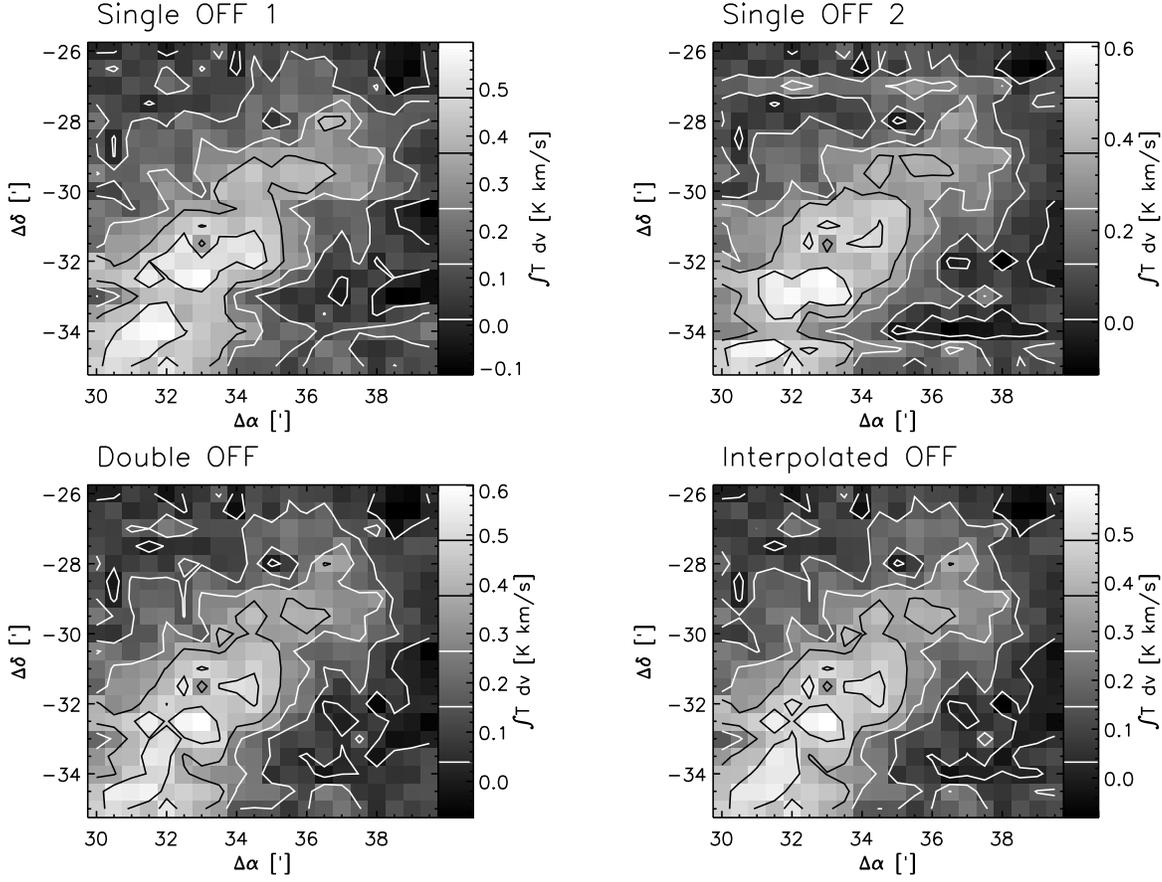}
\caption{\footnotesize Channel map at 5.7~km/s obtained in the different calibration
schemes from the same data as used for the integrated line maps in
Fig. \ref{fig_observedlinemap}.
}
\label{fig_observedlinepeak}
\end{figure*}

\begin{figure}[ht]
\includegraphics[angle=90,width=\columnwidth]{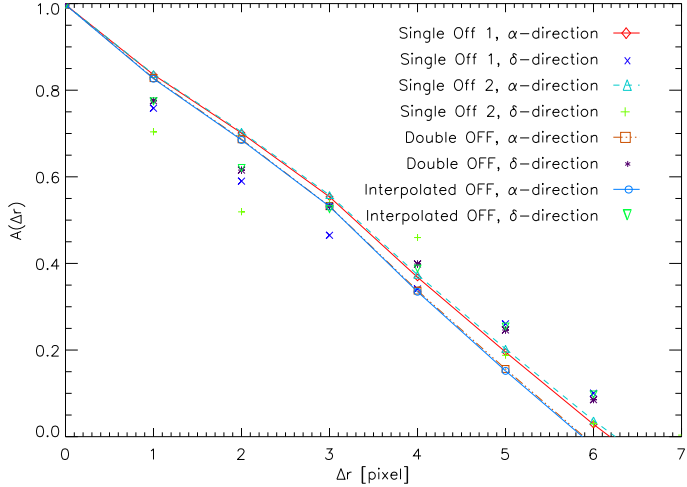}
\caption{\footnotesize Autocorrelation function for the maps from Fig. 
\ref{fig_observedlinepeak} measured in the two perpendicular directions
$\alpha$ and $\delta$. }
\label{fig_acfpeakmap}
\end{figure}

The integrated line map discussed above is strongly susceptible
to drift effects due to the short effective Allan time. To better
study the effect of correlated radiometric noise we can use single 
channel maps where the radiometric noise is higher and the 
relative drift contribution is lower. An ideal observation would give
a {\changed noise of $\sigma(t\sub{tot})\approx 0.23$~K in this map.
With the overheads quoted above this corresponds to 0.30~K 
radiometric noise per point and the maximum
drift noise contribution $\sigma_{C,{\rm drift}}$ computed for the
single-OFF calibration is} only 0.44 $\sigma_{C,{\rm rad.}}$.
Figure \ref{fig_observedlinepeak} shows the maps from the spectrometer 
channel at the peak of the average line profile obtained in the different
calibration schemes. Figure \ref{fig_acfpeakmap} shows the corresponding
autocorrelation functions measuring the anisotropy.  We find, much 
smaller differences between the calibration schemes than in Fig. 
\ref{fig_observedlinemap} but the differences are still dominated by 
drift effects and not by the different level of correlated radiometric
noise. Those lines showing drift signatures in the integrated maps are
the same lines that show the smaller deviations in the channel maps
although the radiometric noise is higher in the channel maps. The
channel maps are much more isotropic than the integrated line maps,
but the maps calibrated by a single OFF still show noticeable stripes.

We can conclude that at least for maps with more that 10 pixels,
the effect of correlated radiometric noise is small so that
the choice of the calibration scheme should be based on the 
capability of correcting the drift of the system.
The application of the different calibration schemes to real 
observations confirms the theoretical considerations that 
the single-OFF calibration is much more susceptible to drift 
effects than the other two calibration schemes so that it should 
be avoided. The example showed no significant advantage of the
interpolated-OFF calibration relative to the double-OFF calibration.
Both schemes lead to a strong reduction of the stripiness of
the resulting maps.

\section{Global optimization}

\subsection{Minimization of the total noise}
\label{sect_optimize}

For any given calibration scheme, the formalism introduced in Sect.
\ref{sect_driftnoisecomputation} can also be used to optimize
the actual timing of the observations. 
We can adapt the scanning speed of the OTF observations to
provide source integration times resulting in a minimum 
uncertainty of the calibrated data.

\begin{figure}[ht]
\includegraphics[angle=90,width=\columnwidth]{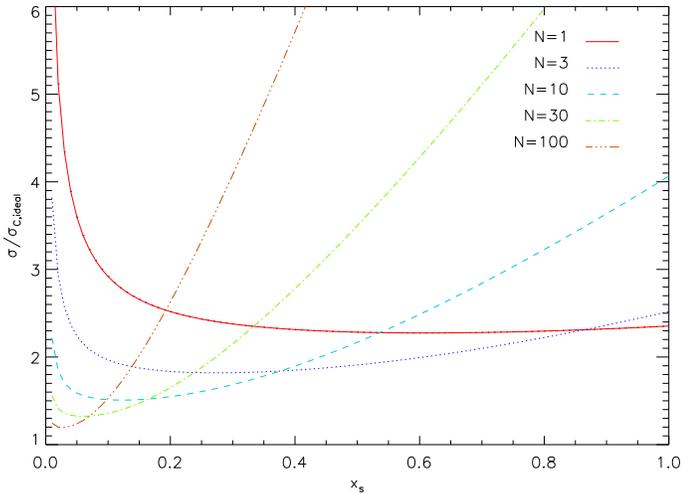}
\caption{\footnotesize RMS of the total data uncertainty from radiometric
and drift noise relative to the noise from an ideal instrument
integrating the same total observing time. The noise is plotted
as a function of the source point integration time
relative to the Allan time for {\changed different numbers of
points per scan, $N$.}
The interpolated-OFF calibration, a spectral index $\alpha=2.5$, 
and a dead time $x\sub{d,1}+x\sub{d,2}=0.25$ were used here.}
\label{fig_n-x-optimize}
\end{figure}

This is demonstrated in Fig. \ref{fig_n-x-optimize} where 
the total data uncertainty is plotted as a function of
the source integration time for different scan lengths
when the interpolated-OFF calibration is used. A spectral
index $\alpha=2.5$ typical of spectroscopic drifts
and a total dead time of 0.25 Allan times were assumed here.
The plot shows the maximum value across the OTF
scans, which is typically reached at the ends of the scan
for short cycles and in the center as soon as drift effects
start to dominate.

For all scan lengths we find a characteristic minimum corresponding
to the optimum source integration time. When using longer
integration times, drift effects start to dominate. At shorter
integration times, the relative overhead from the 
slew to and from the OFF position reduces the observing efficiency
so that the radiometric noise is too high. We also see 
that the relative accuracy of the observations at the optimum
timing grows with the scan size $N$.
The equivalent plot for the double-OFF calibration shows a
steeper increase of the noise when the source integration
time is above its optimum value but minima which are slightly
deeper than the minima shown here.

A special behavior occurs for very shallow spectra ($\alpha \la 0.75$).
They show about the same slope at long integration times for all scan 
lengths so that the curves do not intersect.
This means that very long scans are always favorable even if
the resulting cycle length is much longer than the Allan time.
This can be understood from
the fact that in fluctuation spectra shallower than $1/f$, the noise
is further reduced with increasing integration time, just like in the
familiar case of white noise, but with another slope. The exact
value for the transition to this behavior depends on the dead times
involved but the limit $\alpha = 0.75$ is a good approximation
for most cases.

Independent from the spectral index of the fluctuations we find that
the best observing mode is always given by very long scans with many
points and a very short integration time per point in each
scan. A full observation is obtained from many of these short-time
coverages. This was already shown by \citet{SchiederKramer}. 
Unfortunately,
there are always practical limitations to this approach.
A telescope cannot move arbitrarily fast and the integrated data
cannot be read out and dumped at an infinite data rate. Thus,
the minimum relative integration time $x\sub{min}$ set by the instrument
is a limiting quantity for the optimum OTF timing. 
Moreover, the size of the astronomical source {\changed naturally 
constrains the scan size $N$. Small maps may consist of a limited
number of points $N\sub{max}$ only.} For any given $x\sub{min}$ and
$N\sub{max}$, a plot like 
Fig. \ref{fig_n-x-optimize}, computed for the actual
slewing time $x\sub{d,1}+x\sub{d,2}$, can be used to obtain the 
optimum setup.
In most cases, the solution will still fall at the extreme 
provided by the maximum possible number of points and the minimum
possible integration time.

{\changed At a number of ground-based telescopes (e.g. JCMT, MOPRA,
KOSMA) the implemented OTF mode identifies the OTF scan length $N$ 
with the length of a single line in an OTF map $N\sub{line}$.} 
However, there is no a priori justification for this
identity. {\changed In a more general approach, used e.g. at
IRAM and APEX, multiple lines are combined within one OTF scan.
An even more generalized approach is foreseen for the pointing
mode definitions of the Herschel Space Telescope where an arbitrary
number of points is measured between two OFF measurements, resulting
in scan sizes that may cover also parts of map lines.}
This is partially motivated by the relatively slow
slew to the OFF position {\changed by the telescope. Here, we
consider this most general approach. When combining 
multiple map lines in one OTF scan the turn-around delay between
subsequent lines, $t\sub{turn}$, has} to be taken into account 
when computing the total
noise in the data.  For a point $i$ measured within an OTF
scan of length $N$, the number of turns before and after this point
are $N\sub{turn,1}=(i-1)/N\sub{line}$ and $N\sub{turn,2}=(N-i)/N\sub{line}$,
respectively. In Eq. (\ref{eq_totaluncert}), the total delay before 
the measurement $x\sub{D,1}$
has to be increased by $N\sub{turn,1}x\sub{turn}$, the total delay
after the measurement $x\sub{D,2}$ by $N\sub{turn,2}x\sub{turn}$,
and the total scan length $x\sub{scan} $ by $(N\sub{turn,1}+N\sub{turn,2})
x\sub{turn}$ ,
where $x\sub{turn}$ denotes the turn time relative to the
Allan time, $x\sub{turn}=t\sub{turn}/t\sub{A}$.

\begin{figure}[ht]
\includegraphics[angle=90,width=\columnwidth]{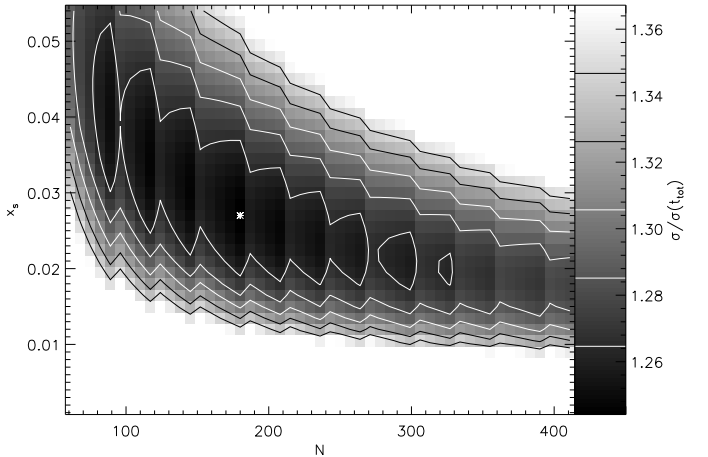}
\caption{\footnotesize Relative RMS of the calibrated data in OTF observations as
a function of the scan length and the integration time per source point
relative to the system Allan time. Values of more than 10\,\% above the minimum
are clipped in the plot. The asterisk marks the optimum setup resulting
in the minimum noise at $N=180$ and $x\sub{s}=0.028$. A line length of 30 points,
an OFF dead time $x\sub{d,1}+x\sub{d,2}=0.6$, a turn time of the telescope
$x\sub{turn}=0.15$, and a spectral index $\alpha=2.5$ were used here.
}
\label{fig_discrete_minimum}
\end{figure}

In this case, the minimum of the relative data uncertainty is no longer
found at the maximum scan length and the minimum possible
integration time because of the increasing overhead for turns
with increasing scan length. The optimization has to be
done by actually evaluating Eq. (\ref{eq_totaluncert}) for
different scan lengths and integration times. This is demonstrated in
Fig. \ref{fig_discrete_minimum} showing the total noise RMS as a function
of source integration time and scan length for a map
where $N\sub{line}=30$ and the relative inter-line overhead is
$x\sub{turn}=0.15$.

The most important feature in this plot is the large extent of the valley.
Within the whole colored part of the plot, the noise RMS only changes
by 10\,\% and even a 2\,\%-contour encloses a factor six in scan length
and a factor two in the integration time. This means that OTF observations
are extremely robust with respect to bad timings. Even in setups far
from the optimum, the noise RMS is typically only enhanced by
some ten percent. This explains why the OTF mode is used very
successfully at many ground-based telescopes without a thorough theoretical
analysis.

The staircase structure of the contours reflects turns
which are added when the scan length exceeds multiples of the
line length. We find several minima with their deepest points always
at integer multiples of the line length. In this example, the 
absolute minimum falls at a scan length of 180 points, but scans 
with 90, 120, 150, or 210 points practically are not
worse. In numerous tests with parameters typical
of different telescopes we found no case with a minimum not
falling at an integer multiple of full lines. Thus we can always complete 
lines in an OTF scan before going to the OFF position and
it is typical that one can combine several lines in an OTF scan.
To optimize actual observations, it is thus sufficient to
evaluate Eq. (\ref{eq_totaluncert}) for scan sizes being
integer multiples of the line length.

With the additional overhead given by the dead time between the
OTF lines, the optimum source integration time is no longer
automatically given by the minimum time allowed by the instrument.
We find an optimum source integration time $x\sub{s}$ between 0.02 and 0.04,
corresponding to a full scan duration of about seven Allan times.
This is a contradiction to the general wisdom, valid for symmetric
observing modes, that the reference cycle period should be shorter
than the Allan time. It can be easily explained by the fact, that 
the Allan time always compares radiometric and drift noise, but
only a small fraction of the full period is used to integrate down
the radiometric noise for any individual map point. However,
this results in a general warning. To make sure that the
proposed optimization scheme is actually valid, the Allan variance
spectrum must not only be known up to a few Allan times, but
it has to be determined over at least the time scale expected for 
the longest OTF observing cycles. This long term spectrum
needs to be used the measure the stability time and to fit the
drift power law to use the optimization formalism derived here.

Eq. (\ref{eq_totaluncert}) can also be used to check the optimum
integration time for the OFF measurement. A rough estimate in 
Sect. \ref{sect_correlated} had shown that for the interpolated-OFF
calibration scheme an OFF integration time of 
$t\sub{OFF}=2/3 \times \sqrt{N} t\sub{s}$ should be sufficient. 
By introducing $q$ as a free parameter
characterizing the relative OFF integration time $t\sub{OFF}=q \sqrt{N}
t\sub{s}$ we can include it in the optimization. {\changed
Searching for the global minimum in the three-dimensional parameter
space spanned by $N$, $x\sub{s}$, and $q$, we find an optimum
value of $q=0.69$ for the parameters used in Fig. \ref{fig_discrete_minimum},
a number that is close to the theoretical value from
Sect. \ref{sect_correlated}. The minimum is, however, very broad
in the $q$-dimension so that} the exact choice of the OFF time
has hardly any influence on the total data uncertainty of the
calibrated data. We find the same kind of robustness as for the scan
lengths.  Varying the model parameters showed that the optimum scan 
length depends strongly on the spectral index of the fluctuations,
but that the optimum $q$-parameter is always close to 0.7.
This value can be used for all observations applying the interpolated-OFF 
or the double-OFF scheme if the full integration time of 
both OFF measurements is used in the calibration.

\begin{table}
\caption{\footnotesize Examples for parameters of an optimum timing in an OTF
map covering 30 points in each line. All values are given for a
spectral resolution of 1~MHz, corresponding to a typical
fluctuation bandwidth of 1.5~MHz.
}
\begin{tabular}[h]{lrr|rrrr}
\hline
observation & $t\sub{A}$ & $\alpha$ & N & $t\sub{s}$ &
$\sigma/\sigma(t\sub{tot})$ & $\sigma^2\sub{drift}/\sigma^2\sub{radio}$ \\
& [s] & & & [s] \\
\hline
\multicolumn{7}{l}{Ground-based telescope: $t_{s,{\rm min}}=1$~s, 
$t\sub{d,1}+t\sub{d,2}=20$~s, $t\sub{turn}=8$~s}\\
\hline
spectroscopic & 80 & 2.5 & 180 & 2 & 1.17 & 0.07 \\
total-power & 8 & 1.5 & 60 & 1 & 1.70 & 0.70 \\
\hline
\multicolumn{7}{l}{HIFI at Herschel: $t_{s,{\rm min}}=1$~s, 
$t\sub{d,1}+t\sub{d,2}=80$~s, $t\sub{turn}=20$~s}\\
\hline
spectroscopic & 150 & 2.5 & 180 & 4 & 1.23 & 0.10 \\
total-power & 10 & 1.5 & 150 & 1 & 2.13 & 1.03 \\
\hline
\end{tabular}
\label{tab_examples}
\end{table}

To provide a feeling for the results, we give in Table \ref{tab_examples}
a few realistic examples. We compare the optimum timing and the 
resulting noise for typical parameters of a ground-based telescope 
and of the HIFI instrument at the Herschel Space Telescope.
{\changed All values are given for a reference resolution of
1~MHz, which is typical of many high-frequency observation,
but not for mm observations of Galactic sources. For such
observations the numbers may only apply to binned or line-integrated 
data.} Ground based telescopes
have the general advantage that they can quickly slew to the
OFF position and store the measured data at a high rate. 
Their big disadvantage is the atmospheric instability resulting
in a relatively short Allan stability time for spectroscopic
and continuum measurements. 
For HIFI observations with the Herschel Space Telescope we expect
a more stable configuration, with an Allan time of about 150\,s
for spectroscopic drifts and of about 10~s for total-power drifts
{\changed (see Sect. \ref{sect_intro})}.
On the other hand,
all telescope slews are relatively slow so that we can estimate
a slewing time of about 40~s when going to an OFF position which is
20$'$ apart from the source.

For spectroscopic observations, {\changed being only sensitive to
relative variations of the sensitivity across the spectrometer,}
we find in both cases an optimum
scan length covering six full lines and $x\sub{s,opt} \approx 
0.025$ which translates into optimum integration times per source 
point of 2~s and 4~s, respectively.
The full range of good observing parameters covers again
scan lengths and integration times differing by up to a factor two
from the optimum values. The drift contribution to the total noise 
is small in both cases, but the ground-based telescope is clearly
superior in terms of the total noise per observing time due to the
shorter dead times. For the Herschel observations the additional
complexity of scanning subsequent lines in opposite directions is
well justified because a limitation of the scan length to the line
length would increase the optimum noise RMS by 9\% corresponding 
to a 19\% loss of observing efficiency.
In {\changed observations for simultaneously determining 
the continuum level, where total-power drifts become relevant,}
the optimum integration time per source
point falls below the 1~s minimum readout time for both
configurations, so that the minimum time provided by the instrument
has to be used. For a 1~s readout and the corresponding optimum 
scan length we find a significant drift contribution with an 
amplitude of 70\,\% of 
the radiometric noise for the ground-based example and 103\,\%
for HIFI at Herschel. This means that the baseline uncertainty
is as large as the radiometric noise contribution. The overall
noise efficiency of the observations is low, relative to an ideal
instrument we obtain 35\,\% for the ground-based 
and 22\,\% for the Herschel continuum observations. Using the
OTF mapping mode for an accurate determination of the
continuum level with heterodyne instruments is therefore 
questionable.

\section{Discussion}
\label{sect_discussion}

One has to keep in mind that OTF observations in general and 
the optimization scheme proposed above in particular have a serious 
drawback. When reusing the calibrated data for purposes which were
not foreseen when planning the observations, by spatial or spectral
rebinning, the relative gain in the noise reduction is always lower than 
for pure radiometric noise. In spatial rebinning the contribution
of the correlated noise from the OFF measurement stays constant
\citep{Beuther}. By extending OTF scans over multiple lines
of a map this effect is in principle even more enhanced. On the
other hand, the total noise contribution from the OFF measurement
drops with increased scan lengths so that the effect of correlated
noise is also somewhat reduced by extending the scans. When applying
the temporal optimization scheme the same becomes true for the
spectral rebinning. The Allan time used to optimize the
observations is determined by the ratio of drift noise and 
radiometric noise. Rebinning spectra to a coarser resolution
only reduces the radiometric noise so that the relative contribution
of the drift noise is enhanced. 

Thus, OTF maps have always a limited use with respect to 
spatial of spectroscopic rebinning. In both
cases artificial structures due to instrumental drifts or due
to the correlated noise are enhanced. Consequently, a very careful
planning of the observations has to be performed. The observer
has to find a compromise between the efficiency of the observations
and their re-usability. The optimization scheme should always be
applied at the level of the coarsest spatial and spectral resolution
that might be used for interpreting the calibrated data. {\changed
The actual data taking can happen at a much higher spatial and 
spectral sampling. For the high sampling the observations will
be less efficient, but the planning then guarantees that
no artifacts will be produced by the foreseen smoothing. For example,
if the observations are taken on a Nyquist sampled grid but a 
smoothing to a half-sampled grid, i.e. a reduction of the
number of independent points by a factor four, is foreseen
the OFF integration time should be approximately doubled
compared to the case where the Nyquist sampling represents the
spatial goal resolution.} Then all artifacts from the observing 
mode are suppressed, however,
at the costs of the observing efficiency. The more precise the
scientific application of the measured data can be specified in terms
of spatial and spectroscopic resolution the better can the actual
observing scheme be adapted to the application resulting in more
efficient observations.

\section{Conclusions}

In most cases mapping observations should follow the scheme
known from OTF maps where the calibration of several source
points uses a common OFF measurement for reference. This is 
far more efficient than all other reference modes. The efficiency
can be further enhanced by combining multiple lines with one OFF measurement.
This introduces, however, correlated noise across the calibrated map
stemming from the common OFF integration. The
impact of this correlated noise can be reduced by using the two 
neighboring  OFF measurements as reference. Their
optimum integration time is approximately $0.7 \sqrt{N} t\sub{s}$.

In most cases the calibration of the source data should follow
the interpolated-OFF scheme where the data from both neighboring
OFF measurements are weighted according to their temporal distance
from the source measurement. This compensates all linear drifts
of the instrument and results in the lowest total uncertainty
of the calibrated data. The single-OFF calibration still used
at several telescopes should be immediately abandoned because of
the strong sensitivity of the calibrated data to drift effects.
For short scans with less than 10 points at a fast telescope
the double-OFF calibration is superior to the interpolated-OFF
calibration. {\changed However, as soon as drift effects may
become important, the robustness of the interpolated-OFF scheme
turns it superior.}

The total uncertainty of the calibrated data consisting of
radiometric noise and drift noise can be computed when the
fluctuation spectrum of instrumental instabilities is known, i.e.
an Allan variance measurement was performed. For a known spectral
goal resolution, the result can
be used to optimize the time line for the actual realization
of the mapping observations. It turns out that the OTF 
observing mode is in general very robust with respect to
non-optimal timings. The scan length and the source integration
time can be varied within a relatively broad range without
increasing the total noise in the calibrated data by more than
a few percent.

The optimization reveals some general relations on conditions
for accurate and efficient mapping observations:
\begin{itemize}
\item{}The efficiency of all mapping modes grows with growing map
size.
\item{}The possibility of fast data readouts is in many cases
essential to minimize the drift contributions.
\item{}In most conditions OTF scans can consist of integer
multiples of complete map lines.
\end{itemize}

The most essential impact on
the data accuracy is provided by the system stability.
All intervals have to be considered relative to the Allan time. The main
prerequisite for any accurate mapping observation is thus a
long instrumental stability, as measured by the Allan time.
Due to the low gain stability of most heterodyne instruments
it turns out that it is often impossible to derive
significant information on the continuum level of
astronomical sources using the OTF or raster mapping modes.
They are always heavily influenced by the instrumental drifts.

Both the general design of the mapping modes with a common
OFF measurement and the temporal optimization limit the
re-usability of the data with respect to spatial or spectroscopic
rebinning. The setup should be optimized with a clear picture
of the resolution requirements set by the scientific goal of
an observation.

\begin{acknowledgements}
I want to thank Rudolf Schieder and Michel {\newchanged P\'erault} for useful
discussions. The research has made use of NASA's Astrophysics Data System
Abstract Service. This work was supported by DLR grant 50 OF 0006.
\end{acknowledgements}

\appendix
\section{Effective beam broadening in OTF maps}

\begin{figure}[ht]
\includegraphics[angle=90,width=\columnwidth]{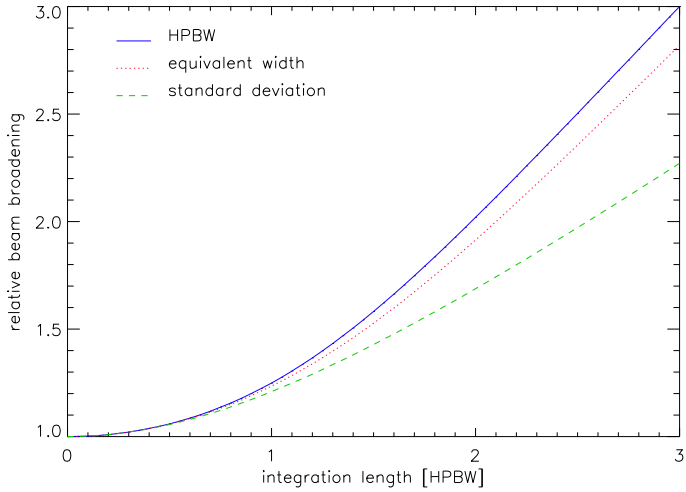}
\caption{\footnotesize Beam broadening due to the scanning motion of the telescope
during integration in OTF observations. The solid line shows the ratio
between the HPBW of the effective beam in scanning direction and the
original Gaussian beam. The dotted line represents the ratio between the
corresponding equivalent widths and the dashed line the ratios
of the standard deviations.}
\label{fig_beamsmear}
\end{figure}

Often OTF observations are not be performed exactly on Nyquist sampling.
Traditionally the difference between a sampling at half the beam width,
HPBW/2, and a full Nyquist sampling is ignored by using the slightly 
coarser sampling. 
For a comparison of different tracers, 
it is moreover useful to map them at the same raster, even when the
observation at different frequencies leads to a slightly different 
sampling with respect to the telescope beam. Thus it is very common
to use samplings deviating from a full Nyquist sampling.
We can compute the quantitative impact of the beam smearing in the
general case by the numerical convolution of a two-dimensional
Gaussian beam profile with a strip function of finite size representing
the motion of the telescope during the integration. The result is
shown in Fig. \ref{fig_beamsmear}. The solid line shows the increase
of the half-power beam width (HPBW) with increased integration length.
We find that the beam broadening goes from the mentioned 4\,\% at 
the Nyquist sampling of 0.42 HPBW to 6\,\% at 0.5 HPBW and to 25\,\%
at 1 HPBW. When using an integration length above 2 HPBW, 
the beam is completely dominated by the strip length. At 2 HPBW, the 
original beam contributes only by 2\,\%.  The beam becomes less and 
less Gaussian. As a measure for the distortion of the beam shape
we have also plotted the ratio between the standard deviation of the 
actual beam and the original beam. It grows much slower than the
HPBW. The beam shape is close to Gaussian for scan lengths below 1 HPBW 
and almost rectangular above 2 HPBW. For an arbitrary beam shape the 
actual resolution is better described by the equivalent width of the 
beam, $\theta\sub{eq}=\int P(\theta) d\theta /P\sub{max}$, \citep{Kraus}. 
This is shown as dotted line in Fig. \ref{fig_beamsmear}. It follows
the curve for the HPBW for integration lengths below 1 HPBW, but
is about 5\,\% lower at long integration lengths.
The beam size perpendicular to the scanning direction is never
affected by the OTF observing scheme. 

\section{Signal correlations from power-law autocorrelation functions}

{\changed Noise functions exhibiting a $1/f^{\alpha}$ power spectrum 
($0< \alpha \le 3, \alpha \ne 1$) are characterized by even power-law 
autocorrelation functions
\begin{equation}
\gamma(\tau)=g_0 -g_{\alpha} |\tau|^{\alpha-1}\;.
\label{eq_corrappx}
\end{equation}

The average correlation of the signal measured over a period $t_1$ with
the signal measured over $t_2$ separated by an arbitrary delay $d$
can be computed as
\begin{eqnarray}
\left\langle c_1(t)c_2(t) \right\rangle_t &=&
{1 \over t\sub{1} t \sub{2}} 
\int_{0}^{t\sub{1}} \!\!d\tau' \int_{0}^{t_2} \!\!d\tau''
\gamma(\tau'-\tau''+d+t\sub{s}) \\
&=& {1 \over t\sub{1} t \sub{2}} 
\int_{0}^{t\sub{1}} \!\!d\tau' \int_{0}^{t_2} \!\!d\tau''
\left(g_0-|\tau'-\tau''+d+t\sub{s}|^{\alpha-1}\right). \nonumber
\end{eqnarray}

We can always chose $t_1$ to start before or at the same time
as $t_2$, so that $t_1+d \ge 0$. Then we obtain
\nonumber\\
\begin{eqnarray}
\left\langle c_1(t)c_2(t) \right\rangle_t &=&g_0-
{1 \over t\sub{1} t \sub{2}} 
\int_{0}^{t\sub{1}} d\tau'
{g_{\alpha} \over \alpha} \left[
|\tau'+d|^\alpha + (\tau'+d+t\sub{1})^\alpha \right] \nonumber\\
&=&g_0-{g_{\alpha} \over \alpha(\alpha+1) t\sub{1} t \sub{2}} \left[
(t\sub{1}+t\sub{2}+d)^{\alpha+1}
- (t\sub{1}+d)^{\alpha+1} \right.\nonumber \\
&& \hspace{2.7cm} \left.
- |t\sub{2}+d|^{\alpha+1}
+ |d|^{\alpha+1} \right]
\end{eqnarray}

The correlation between the signals measured in $t_1$ and $t_2$
decays basically with the total time spanned by the two integrations
and the delay between them (first term in brackets), reduced
by some corrections for the finite integration times.

One special case is the auto-correlation over the same time interval,
i.e. $t_1=t_2=-d$. Then we obtain the simple expression
\begin{equation}
\left\langle c_1(t)^2 \right\rangle_t =
g_0-{2 g_{\alpha} t\sub{1}^{\alpha-1} \over \alpha(\alpha+1)}
\;.
\label{eq_alpharadio}
\end{equation}
Another important case is the correlation measured in Allan
variance measurements with $t_1=t_2$, $d=0$. The correlation
between two adjacent intervals of the same length is
\begin{equation}
\left\langle c_1(t)c_2(t) \right\rangle_t =
g_0-{2 g_{\alpha} (2^\alpha -1) t\sub{1}^{\alpha-1} \over \alpha(\alpha+1)}
\;.
\end{equation}

For the case of white noise, $\alpha=0$, Eq. (\ref{eq_correlation})
does not hold, but the correlation function represents a Dirac
$\delta$-function, $\gamma(\tau)=g_0 \delta(|\tau|)$. We
obtain
\begin{eqnarray}
\left\langle c_1(t)c_2(t) \right\rangle_t &=&
{1 \over t\sub{1} t \sub{2}} 
\int_{0}^{t\sub{1}} \!\!d\tau' \int_{0}^{t_2} \!\!d\tau''
\gamma_0 \delta(\tau'-\tau''+d+t\sub{s}) \\
&=& {g_0 \over t\sub{1} t \sub{2}} 
\int_{0}^{t\sub{1}} \!\!d\tau' \left\{
\begin{array}{ll} 1 &\mbox{if $0 \le \tau' \le -d$}\\ 0 & \mbox{otherwise}
\end{array}\right. \nonumber \\
&=&\left\{
\begin{array}{ll} {- g_0 d / (t\sub{1} t \sub{2})} &\mbox{if $d<0$}\\
0 & \mbox{otherwise}
\end{array}\right.
\end{eqnarray}
This means that a noise correlation occurs only within the period
of overlap of the two measurements given by the negative delay, $-d$.
For all positive or zero delays, like in the Allan variance measurement,
the correlation vanishes.
The special case of the auto-correlation over the same time interval,
leads to the well known radiometric noise behavior
$\left\langle c_1(t)^2 \right\rangle_t = g_0/t_1$.}

\section{Raster map observations}

Raster map observations differ from the OTF observations discussed
in the main part of this paper by pointing individually at the different
source map points instead of continuously scanning over the source.
This has two effects. First, the effective beam of the observation is
always equal to the actual telescope beam.  It does not suffer from 
the beam broadening discussed for OTF maps in Sect. \ref{sect_genscheme}. 
Second, it introduces dead times between the observation of different
points of a map. The observation of each source point is characterized
by two time constants here, the source integration time $t\sub{s}$
and the slew time to the next map position $t\sub{m}$.  For all
points in the map, except for the last point of a scan, the
total time needed for the measurement is given by
$t\sub{s,tot}=t\sub{s}+t\sub{m}$. No additional turn time between two map lines
is required. If we redefine the slew time to the OFF position
as $t\sub{d,2}'=t\sub{d,2}-t\sub{m}$ we can use all equations
derived above for the calibration and the noise estimate in 
the OTF mode by using $t\sub{s}$ whenever the
integration time counts and $t\sub{s,tot}$ whenever delays enter.

In particular the interpolation measure $l$ derived in Sect.
\ref{sect_calibration} (Eq. \ref{eq_interpolmeasure}) turns into
\begin{equation}
l={{t\sub{R}/2+t\sub{d,1}+(i-1/2) t\sub{s,tot}} \over {t\sub{R}
+t\sub{d,1}+t\sub{d,2}'+N t\sub{s,tot}}}
\end{equation}

For the estimate of the total noise in the data Eq. (\ref{eq_totaluncert})
can still be used when the total delays include the additional
slew times, i.e.
\begin{eqnarray}
x\sub{D,1}&=&x\sub{d,1}+(i-1)x\sub{s,tot}\nonumber\\
x\sub{D,2}&=&x\sub{d,2}+(N-i)x\sub{s,tot}=x\sub{d,2}'+(N-i)x\sub{s,tot}+x\sub{m}\nonumber\\
x\sub{scan}&=&x\sub{d,1}+Nx\sub{s,tot}+x\sub{d,2}'
\end{eqnarray}

The resulting general behavior corresponds to an OTF map with
very long delays before and after the lines. The corresponding optimum
timing may consist of scan lengths which are shorter than the
line lengths but there are no qualitative differences to the 
properties discussed for OTF observations.

\end{document}